\documentclass[11pt]{article}
\newtheorem{lemma}{Lemma}
\newtheorem{definition}{Definition}
\usepackage{cite}
\usepackage{amsmath}
\usepackage{amssymb}
\usepackage{psfig}
\usepackage{epsfig}
\usepackage{color}

\newcommand{\bigsqcap}{\mbox{$\Large \sqcap$}}

\newcommand{\true}{\mbox{\em true}}
\newcommand{\false}{\mbox{\em false}}

\newcommand{\hst}{\\\hspace*{4mm}}
\newcommand{\hstt}{\\\hspace*{8mm}}
\newcommand{\hsttt}{\\\hspace*{12mm}}
\newcommand{\hstttt}{\\\hspace*{16mm}}

\newcommand{\pf}{\noindent\mbox{\bf Proof : }}

\newcommand{\procbegin}{\vspace*{3mm}\hrule\vspace{2mm}\noindent}
\newcommand{\procend}{\vspace*{2mm}\hrule\vspace{3mm}}

\def\qed{\ifmmode\|\else{\unskip\nobreak\hfil
\penalty50\hskip1em\null\nobreak\hfil$\blacksquare$
\parfillskip=0pt\finalhyphendemerits=0\endgraf}\fi}

\newenvironment{list1}{\begin{list}{$\bullet$}
{\topsep 0 pt \parsep 0 pt \partopsep 0 pt \itemsep 0 pt}}{\end{list}}

\newcounter{cabbage1}

\newcounter{cabbage2}

\newcounter{cabbage3}

\newcounter{bean1}

\newcounter{bean2}

\newcounter{bean3}

\newcounter{bean4}

\newcounter{bean5}

\newcounter{bean6}

\begin{document}


\title{Symbolic Parametric Analysis of Linear Hybrid Systems 
with BDD-like Data-Structures\thanks{The
work is partially supported by NSC, Taiwan, ROC under
grants NSC 90-2213-E-002-131, NSC 90-2213-E-002-132}\thanks{A small part of the materials
was presented in an invited speech at AVIS, April 2003, Warsaw, Poland.}\thanks{This article
is also available at ACM CoRR (Computing Research Repository, http://www.acm.org/corr/),
user:cs.DS/0306113.
}\thanks{We would like to express special thanks to the TReX team who have helped 
us to use TReX in the experiment.  
Especially, since we haven't purchased the library package used in TReX for backward analysis, 
Mihaela Sighireanu and Aurore Collomb-Annichini 
has kindly collected the performance data of 
TReX in backward analysis for us.  
}
}

\author{Farn Wang \\ 
%
Dept. of Electrical Engineering,
National Taiwan University\\
1, Sec. 4, Roosevelt Rd., Taipei, Taiwan 106, ROC;\\
+886-2-23635251 ext. 435; FAX:+886-2-23671909; \\
farn@cc.ee.ntu.edu.tw;
http://cc.ee.ntu.edu.tw/\~{ }farn\\[2mm]
Model-checker/simulator {\tt red} 5.0 will be available at
http://cc.ee.ntu.edu.tw/\~{ }val
}

\date{}

\maketitle

\begin{abstract}
We use dense evaluation ordering to define HRD (Hybrid-Restriction Diagram),
a new BDD-like data-structure for the representation and manipulation of
state-spaces of linear hybrid automata.
We present and discuss various manipulation algorithms for HRD, including
the basic set-oriented operations, weakest precondition calculation, and normalization.
We implemented the ideas and experimented to see their performance.
Finally, we have also developed a pruning technique for state-space exploration based
on parameter valuation space characterization.
The technique showed good promise in our experiment.
\end{abstract}


\section{Introduction}\label{sec:intro}

Symbolic analysis of {\em linear hybrid automata} (LHA)\cite{AHH93,ACHHHNOSS95}
can generate a symbolic characterization of the reachable state-space of the LHA.
When {\em static parameters} (system variables whose values are decided before run-time and 
never changed in run-time) are used in LHAs, such symbolic characterizations may
shed important feedback information to engineers.
For example, we may use such symbolic characterizations to choose proper parameter values
to avoid from unsafe system designs.
Unfortunately, LHA systems are extremely complex and not subject to algorithmic
analysis\cite{AHV93}.
Thus in real-world applications, it is very important to use every measure to
enhance the efficiency of LHA parametric analysis.

In this work, we extend BDD-like data-structures\cite{BCMDH90,Bry86} for the representation and manipulation
of LHA state-spaces.
BDD-like data-structures have the advantage of data-sharing in both representation and
manipulation and have shown great success in VLSI verification industry.
One of the major difficulties to use BDD-like data-structures to
analyze LHAs comes from the unboundedness of the dense variable value ranges and
the unboundedness of linear constraints.
To explain one of the major contribution of this work, we need to discuss the
following issue first.
In the research of BDD-like data-structures,
there are two classes of variables: {\em system variables} and {\em decision atoms}\cite{Wang03}.
System variables are those used in the input behavior descriptions.
Decision atoms are those labeled on each BDD nodes.
For discrete systems, these two classes are the same,
that is, decision atoms are exactly those system variables.
But for dense-time systems, decision atoms can be different from state varaibles.
For example, in CDD\cite{BLPWW99} and CRD\cite{Wang03},
decision atoms are of the form $x-x'$ where $x$ and $x'$ are system variables of
type clock.
Previous work on BDD-like data-structures are based on the
assumption that decision atom domains are of finite sizes.
Thus we need new techniques to extend BDD-like data-structures to
represent and manipulate state-spaces of LHAs.
Our innovations include using constraints like $-3A+x-4y$ (where $A,x,y$ are dense variables),
as the decision atoms and using total dense orderings among these atoms.
In this way, we devised HRD (Hybrid-Restriction Diagram) and successfully extend
BDD-technology to models with unbounded domains of decision atoms.

In total, we defined three total dense-orderings for HRD constriants (section~\ref{sec.vorderings}).
We also present algorithms for set-oriented operations (section~\ref{sec.set}) and
symbolic weakest precondition calculation (section~\ref{sec.wpc}),
procedures for symbolic parametric analysis (section~\ref{sec.wpc}), and
discuss our implementation of
symbolic convex polyhedra representation normalization (section~\ref{sec.norm}).
Especially, in the presentation of our previous work of BDD-like data-structures
for timed automata, people usually asked for presentation of our algorithms for weakest
precondition construction.
In this paper, we endeavored to make a concise presentation.

We have also developed a techique for fast parametric analysis of LHA (section~\ref{sec.pspsc}).
The technique prunes state-space exploration based on static parameter space characterization.
The technique gives us very good performance.
Desirably, this technique does not sacrifice the precision of parametric analysis. 
Especially, for one benchmark, the state-space exploration does not converge  
without this technique!  
To our knowledge, nobody else has come up with a similar technique.
Finally, we have implemented our ideas in our tool {\tt red} 5.0 and reported our experiments
to see how the three dense-orderings perform and
how our implementation performs in comparison 
with HyTech 2.4.5\cite{HHWt95} and TReX 1.3\cite{AAB00,ABS01}.

\section{Related work \label{sec.relwork}}

Many modern model-checkers\cite{PL00,Wang03,Yovine97} for timed automata\cite{AD89}
are built around symbolic manipulation
procedures\cite{HNSY92,AHH93} of {\em zones},
which means behaviorally equivalent
convex state spaces of timed automata.
The most popular data-structure for zones is DBM\cite{Dill89}, which is
a two dimensional matrix recording differences
between pairs of clocks and nothing BDD-like.

As far as we know, the first paper that discusses how to use BDD to 
encode zones is by Wang, Mok, and Emerson in 1993\cite{WME93}.
They discussed how to use BDD with decision atoms like $x_i+c\leq x_j+d$ 
to model-check timed automata.   
Here $c$ and $d$ are timing constants with magnitude $\leq C_A$.  
However, they did not report implementation and experiments.
In the last several years, people have explored in this approach 
in the hope to duplicate the
success of BDD techniques\cite{BCMDH90,Bry86} in hardware verification
for the verification of timed automata
\cite{ABKMPR97,Balarin96,BLPWW99,MLAH99a,MLAH99b,Wang00a,Wang00b,Wang01a,Wang01b,Wang03}.

For parametric analysis, 
Annichini et al have extended DBM to PDBM 
for parametric analysis of timed automata\cite{AAB00,ABS01}
and implemented a tool called {\em TReX}, which also supports 
verification with lossy channels.  
Due to the differences in their target systems, it can be difficult 
to directly compare the performances of TReX and our implementation {\tt red} 5.0. 
For example, TReX only allows for clocks while {\tt red} 5.0 
allows for dense variables with rate intervals.  
To construct time-progress weakest preconditions (or strongest postcondition in 
forward analysis) for systems with dense variable rate intervals, 
{\tt red} 5.0 needs to use one $\delta$-variable for each dense variables 
and significantly increase the number of decision atoms involving $\delta$-variables.  
According to the new formulation of time-prorgress weakest precondition algorithm 
in \cite{Wang03}, for systems with only clocks, 
no literals involving $\delta$-variables ever need to be generated. 
Thus the complexity for the algorithm used in {\tt red} 5.0  
is relatively higher than those used in TReX.  
On the other hand, TReX may have tuned its performance for the verification 
of lossy channel systems.  

For LHAs, people also used convex subspaces, called {\em convex polyhedra},
as basic unit for symbolic manipulation.
A convex polyhedron characterizes a state-space of an LHA and
can be symbolically represented by a set of
constraints like $a_1x_1+\ldots+a_nx_n\sim c$ \cite{ACHH93,AHH93,ACHHHNOSS95}.
Two commonly used representations for convex polyhedra in HyTech are
(1) polyhedras and (2) frames in dense state-space\cite{HHWt95}.
These two representations neither are BDD-like nor can represent concave state-spaces.
Data-sharing among convex polyhedra is difficult.

\section{Parametric analysis of linear hybrid automata (LHA)\label{sec.lha}}

A {\em linear hybrid automata (LHA)}\cite{AHH93}
is a finite-state automaton equipped with
a finite set of dense variables which can hold real-values.
At any moment, the LHA can stay in only one {\em mode} (or {\em control location}).
In its operation, one of the transitions can be triggered
when the corresponding triggering condition is satisfied.
Upon being triggered, the LHA instantaneously transits from one
mode to another and sets some dense variables to values in certain ranges.
In between transitions, all dense variables increase their readings at rates
determined by the current mode.

For convenience, given a set $Q$ of modes and a set $X$ of dense variables,
we use $P(Q,X)$ as
the set of all Boolean combinations of atoms of the forms $q$ and
$\sum a_i x_i\sim c$,  where $q\in Q$, $a_i$ are integers constants,
$x_i\in X$, ``$\sim$'' is one of
$\leq, <,=,>,\geq$, and $c$ is a rational constant.

We also let $\Theta$ be the set of rational intervals like
$\langle d,d'\rangle$ where
'$\langle$' is either '$[$' or '$($';
'$\rangle$' is either '$]$' or '$)$'; and
$d,d'$ are $-\infty,\infty$, or rational numbers.

{\definition \underline{\bf linear hybrid automata (LHA)}}
An LHA $A$ is a tuple \linebreak 
$\langle X, Q, I, \mu, \gamma, E, \tau, \pi\rangle$
with the following restrictions.
$X$ is the set of dense variables.
$Q$ is the set of  modes.
$I\in P(Q,X)$ is the initial condition.
$\mu:Q\mapsto P(\emptyset,X)$ defines the invariance condition of each mode.
$\gamma:(Q\times X)\mapsto \Theta$ defines the rate intervals of dense variables.
$E\subseteq Q\times Q$ is the set of transitions.
$\tau:E\mapsto P(\emptyset,X)$ defines the triggering condition of transitions.
$\pi$ is a partial function from $E\times X$ to $\Theta$ that
defines the interval assignments to dense variables
at each transition.
If $\pi(e,x)$ is undefined, $x$ is not assigned a value in transition $e$;
otherwise, $x$ is nondeterministically assigned a finite rational value in
$\pi(e,x)$ in transition $e$.
\qed
\vspace*{2mm}

\begin{figure}[t]
\begin{center}
\begin{picture}(0,0)%
\includegraphics{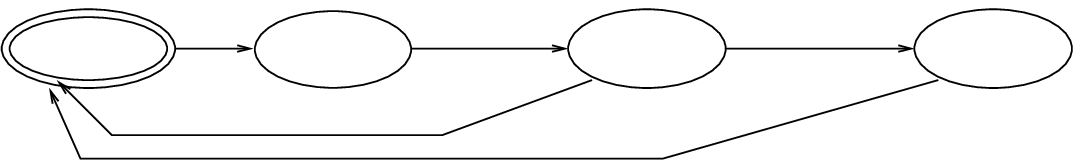}%
\end{picture}%
\setlength{\unitlength}{3947sp}%
\begingroup\makeatletter\ifx\SetFigFont\undefined%
\gdef\SetFigFont#1#2#3#4#5{%
  \reset@font\fontsize{#1}{#2pt}%
  \fontfamily{#3}\fontseries{#4}\fontshape{#5}%
  \selectfont}%
\fi\endgroup%
\begin{picture}(5153,737)(143,-940)
\put(453,-362){\makebox(0,0)[lb]{\smash{\SetFigFont{6}{7.2}{\rmdefault}{\mddefault}{\updefault}{\color[rgb]{0,0,0}$q_0$}%
}}}
\put(1662,-324){\makebox(0,0)[lb]{\smash{\SetFigFont{6}{7.2}{\rmdefault}{\mddefault}{\updefault}{\color[rgb]{0,0,0}$q_1$}%
}}}
\put(1964,-777){\makebox(0,0)[lb]{\smash{\SetFigFont{6}{7.2}{\rmdefault}{\mddefault}{\updefault}{\color[rgb]{0,0,0}$L\neq P$}%
}}}
\put(3173,-324){\makebox(0,0)[lb]{\smash{\SetFigFont{6}{7.2}{\rmdefault}{\mddefault}{\updefault}{\color[rgb]{0,0,0}$q_2$}%
}}}
\put(2153,-362){\makebox(0,0)[lb]{\smash{\SetFigFont{6}{7.2}{\rmdefault}{\mddefault}{\updefault}{\color[rgb]{0,0,0}$x<\alpha$}%
}}}
\put(2153,-513){\makebox(0,0)[lb]{\smash{\SetFigFont{6}{7.2}{\rmdefault}{\mddefault}{\updefault}{\color[rgb]{0,0,0}$x:=0; L:=P;$ }%
}}}
\put(3664,-362){\makebox(0,0)[lb]{\smash{\SetFigFont{6}{7.2}{\rmdefault}{\mddefault}{\updefault}{\color[rgb]{0,0,0}$x\geq\beta\wedge L=P$}%
}}}
\put(1020,-362){\makebox(0,0)[lb]{\smash{\SetFigFont{6}{7.2}{\rmdefault}{\mddefault}{\updefault}{\color[rgb]{0,0,0}$L=0$}%
}}}
\put(1020,-513){\makebox(0,0)[lb]{\smash{\SetFigFont{6}{7.2}{\rmdefault}{\mddefault}{\updefault}{\color[rgb]{0,0,0}$x:=0;$ }%
}}}
\put(4873,-324){\makebox(0,0)[lb]{\smash{\SetFigFont{6}{7.2}{\rmdefault}{\mddefault}{\updefault}{\color[rgb]{0,0,0}$q_3$}%
}}}
\put(3022,-890){\makebox(0,0)[lb]{\smash{\SetFigFont{6}{7.2}{\rmdefault}{\mddefault}{\updefault}{\color[rgb]{0,0,0}$L:=0;$ }%
}}}
\put(227,-475){\makebox(0,0)[lb]{\smash{\SetFigFont{6}{7.2}{\rmdefault}{\mddefault}{\updefault}{\color[rgb]{0,0,0}$\dot{x}\in[4/5,1]$}%
}}}
\put(1435,-475){\makebox(0,0)[lb]{\smash{\SetFigFont{6}{7.2}{\rmdefault}{\mddefault}{\updefault}{\color[rgb]{0,0,0}$\dot{x}\in[4/5,1]$}%
}}}
\put(2946,-475){\makebox(0,0)[lb]{\smash{\SetFigFont{6}{7.2}{\rmdefault}{\mddefault}{\updefault}{\color[rgb]{0,0,0}$\dot{x}\in[4/5,1]$}%
}}}
\put(4608,-475){\makebox(0,0)[lb]{\smash{\SetFigFont{6}{7.2}{\rmdefault}{\mddefault}{\updefault}{\color[rgb]{0,0,0}$\dot{x}\in[4/5,1]$}%
}}}
\end{picture}
\end{center}
\caption{Fischer's timed mutual exclusion algorithm in LHA}
\label{fig.fischerc}
\end{figure}

In figure~\ref{fig.fischerc}, we have drawn a version of
the Fishcer's mutual exclusion algorithm for a process.
There are two static parameters $\alpha$ and $\beta$ that controls the
behavior of the processes.
In each mode, local clock $x$ increases its reading according to
a rate in $[4/5,1]$.
The rate interval in each mode can be different.

A {\em valuation} of a set is a mapping from the set to another set.
Given an $\eta\in P(Q,X)$ and a valuation $\nu$ of $X$, we say
$\nu$ {\em satisfies} $\eta$, in symbols $\nu\models\eta$,
iff it is the case that
when the variables in $\eta$ are interpreted according to $\nu$,
$\eta$ will be evaluated $\true$.
{\definition \underline{\bf states}}
A {\em state} $\nu$ of $A=\langle X, Q, I, \mu, \gamma, E, \tau, \pi\rangle$
is a valuation of $X\cup Q$ s.t.
\begin{list1}
\item there is a unique $q\in Q$ such that $\nu(q)=\true$ and
	for all $q'\neq q, \nu(q')=\false$;
\item for each $x\in X$, $\nu(x)\in {\cal R}$
	(the set of reals)
	and $\forall q\in Q, \nu(q)\Rightarrow \nu\models\mu(q)$.
\end{list1}
Given state $\nu$ and $q\in Q$ such that $\nu(q)=\true$,
we call $q$ the mode of $\nu$, in symbols $\nu^Q$.
\qed
For any $t\in {\cal R}^+$ (the set of nonnegative reals), $\nu\stackrel{t}{\leadsto}\nu'$ iff
we can go from $\nu$ to $\nu'$ merely by the passage of $t$ time units.
Formally speaking, $\nu\stackrel{t}{\leadsto}\nu'$ is true iff
$\nu'$ is a state identical to $\nu$
except that for every $x\in X$ with $\gamma(\nu^Q,x)=\langle d,d'\rangle$,
$\nu'(x)\in \langle \nu(x)+t\cdot d,\nu(x)+t\cdot d'\rangle$.

For a transition $e\in E$, $\nu\stackrel{e}{\leadsto}\nu'$ iff
we can go from $\nu$ to $\nu'$ with discrete transition $e=(q,q')$.
Formally speaking, $\nu\stackrel{e}{\leadsto}\nu'$ is true iff
$\nu^Q=q$, $\nu\models \mu(q)\wedge \tau(e)$, and
$\nu'$ is identical to $\nu$ except that
\begin{list1}
\item $\nu'^Q=q'$ and $\nu'\models \mu(q')$; and
\item for each $x\in X$, if $\pi(e,x)$ is defined, 
	$\nu'(x)\in \pi(e,x)$; otherwise, $\nu'(x)=\nu(x)$;
\end{list1}

{\definition \underline{\bf runs}}
Given an LHA $A=\langle X, Q, I, \mu, \gamma, E, \tau, \pi\rangle$,
a {\em run} is an infinite sequence of pairs
$(\phi_0,t_0)(\phi_1,t_1)\ldots(\phi_k,t_k)\ldots\ldots$
such that $t_0 t_1 \ldots t_k\ldots\ldots$
is a monotonically increasing real-number (time)
divergent sequence
and for all $k\geq 0$,
\begin{list1}
\item $\phi_k$ is a mapping from $[t_k,t_{k+1}]$ to states, and
\item {\em time-progress is continuous:} that is, \\
	for each $t_k\leq t\leq t'\leq t_{k+1}$,
	$\phi_k(t)\stackrel{t'-t}{\leadsto}\phi_k(t')$; and
\item {\em invariance conditions are preserved in each interval:} that is,\\
	for all $t_k\leq t\leq t_{k+1}$,
	$\phi_k(t)\models\mu(\phi_k(t)^Q)$; and
\item either {\em no transition happens at time $t_{k+1}$}, that is,
	$\phi_k(t_{k+1})^Q=\phi_{k+1}(t_{k+1})^Q$;
	or
	{\em a transition $e$ happens at $t_{k+1}$}, that is,
	$\phi_k(t_{k+1})\stackrel{e}{\leadsto}\phi_{k+1}(t_{k+1})$.
\qed
\end{list1}

A run $\rho=(\phi_0,t_0)(\phi_1,t_1)\ldots(\phi_k,t_k)\ldots\ldots$ is {\em safe} w.r.t.
a safety state-predicate $\eta$, in symbols $\rho\models \eta$,
iff for all $k\geq 0$ and $t\in [t_k,t_{k+1}]$, $\phi_k(t)\models\eta$.
A dense variable $x$ in an LHA is a {\em static parameter} iff
its rate is always zero in all modes.
Suppose $H$ is the set of static parameters in $X$ of LHA $A$.
A {\em static parameter valuation} $\cal H$
of a run $(\phi_0,t_0)(\phi_1,t_1)\ldots(\phi_k,t_k)\ldots\ldots$
is a mapping from $H$ to reals such that $\cal H$ is consistent with every state along $\rho$,
i.e., $\forall x\in H\forall k\geq 0(\phi_k(t_k)(x)={\cal H}(x))$.
$\cal H$ is a {\em parametric solution} to $A$ and $\eta$ iff
for all runs $\rho$ with static parameter valuation $\cal H$,
$\rho\models\eta$.

Our verification framework is called {\em parametric safety analysis problem}.
A parametric safety analysis problem instance, $\mbox{PSA}(A,\eta)$ in notations,
consists of an LHA $A$ and a safety state-predicate $\eta\in P(Q,X)$.
Such a problem instance asks for a symbolic characterization of all parametric solutions
to $A$ and $\eta$.
The general parametric safety analysis problem is undecidable.

\section{Convex polyhedra\label{sec.cpolyhedra}}

Given a set $X=\{x_1,\ldots, x_n\}$ of dense system variables,
an {\em LH-expression (linear hybrid expression)}
is an expression like $a_1x_1+\ldots+a_nx_n$ where $a_1,\ldots,a_n$ are integer constants.
It is {\em normalized} iff the gcd of nonzero coefficients in $\{a_1,\ldots,a_n\}$ is 1,
i.e., $\gcd\{a_i\mid 1\leq i\leq n; a_i\neq 0\}=1$.
From now on, we shall assume that all given LH-expressions are normalized.

An {\em LH-upperbound} is either $(<,\infty)$ or
a pair like $(\sim, c)$ where $\sim\in\{"<", "\leq"\}$ and $c$ is a rational number.
There is a natural ordering $\sqsubset$ among the LH-upperbounds.
That is for any two $(\sim,c)$ and $(\sim',c')$, $(\sim,c)\sqsubset(\sim',c')$ iff
$c<c'$ or $(c=c' \wedge\sim="<"\wedge \sim'="\leq")$.
Intuitively, if $(\sim,c)\sqsubset (\sim',c')$, then $(\sim,c)$ is more restrictive than
$(\sim',c')$.

An {\em LH-constraint} is a pair of an LH-expression and an LH-upperbound.
Given an LH-expression $\sum_ia_ix_i$ and an LH-upperbound $(\sim c)$,
we shall naturally write the corresponding LH-constraint as
$\sum_ia_ix_i\sim c$.
A {\em convex polyhedron} is symbolically represented by a conjunction of LH-constraints
and means a behaviorally equivalent state subspace of an LHA.
Formally, a convex polyhedron $\zeta$ can be defined as a mapping from the set of
LH-expressions to the set of LH-upperbounds.
Alternatively, we may also represent a convex polyhedron $\zeta$
as the set $\{\sum_ia_ix_i\sim c\mid \zeta(\sum_ia_ix_i)=(\sim, c)\}$.
We shall use the two equivalent notations flexibly as we see fit.
With respect to a given $X$, the set of all LH-expressions and the set of convex polyhedra
are both infinite.

\section{HRD (Hybrid-Restriction Diagram) \label{sec.HRD}}

To construct BDD-like data-structures, three fundamental issues have to be solved.
The first is the domain of the decision atoms;
the second is the range of the arc labels from BDD nodes;
and the third is the evaluation ordering among the decision atoms.
For modularity of presentation, we shall leave the discussion of the
evaluation orderings to section~\ref{sec.vorderings}.
In this section, we shall assume that we are given a decision atom evaluation ordering.

\begin{figure}[t]
\begin{center}
\begin{picture}(0,0)%
\includegraphics{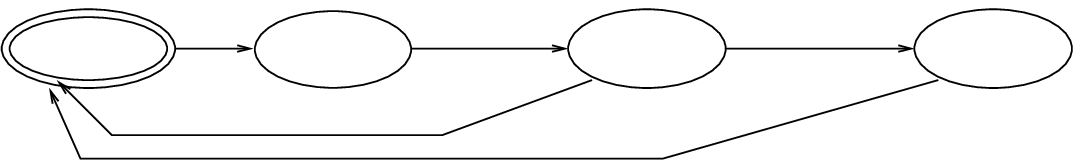}%
\end{picture}%
\setlength{\unitlength}{3947sp}%
\begingroup\makeatletter\ifx\SetFigFont\undefined%
\gdef\SetFigFont#1#2#3#4#5{%
  \reset@font\fontsize{#1}{#2pt}%
  \fontfamily{#3}\fontseries{#4}\fontshape{#5}%
  \selectfont}%
\fi\endgroup%
\begin{picture}(5153,737)(143,-940)
\put(453,-362){\makebox(0,0)[lb]{\smash{\SetFigFont{6}{7.2}{\rmdefault}{\mddefault}{\updefault}{\color[rgb]{0,0,0}$q_0$}%
}}}
\put(1662,-324){\makebox(0,0)[lb]{\smash{\SetFigFont{6}{7.2}{\rmdefault}{\mddefault}{\updefault}{\color[rgb]{0,0,0}$q_1$}%
}}}
\put(1964,-777){\makebox(0,0)[lb]{\smash{\SetFigFont{6}{7.2}{\rmdefault}{\mddefault}{\updefault}{\color[rgb]{0,0,0}$L\neq P$}%
}}}
\put(3173,-324){\makebox(0,0)[lb]{\smash{\SetFigFont{6}{7.2}{\rmdefault}{\mddefault}{\updefault}{\color[rgb]{0,0,0}$q_2$}%
}}}
\put(2153,-362){\makebox(0,0)[lb]{\smash{\SetFigFont{6}{7.2}{\rmdefault}{\mddefault}{\updefault}{\color[rgb]{0,0,0}$x<\alpha$}%
}}}
\put(2153,-513){\makebox(0,0)[lb]{\smash{\SetFigFont{6}{7.2}{\rmdefault}{\mddefault}{\updefault}{\color[rgb]{0,0,0}$x:=0; L:=P;$ }%
}}}
\put(3664,-362){\makebox(0,0)[lb]{\smash{\SetFigFont{6}{7.2}{\rmdefault}{\mddefault}{\updefault}{\color[rgb]{0,0,0}$x\geq\beta\wedge L=P$}%
}}}
\put(1020,-362){\makebox(0,0)[lb]{\smash{\SetFigFont{6}{7.2}{\rmdefault}{\mddefault}{\updefault}{\color[rgb]{0,0,0}$L=0$}%
}}}
\put(1020,-513){\makebox(0,0)[lb]{\smash{\SetFigFont{6}{7.2}{\rmdefault}{\mddefault}{\updefault}{\color[rgb]{0,0,0}$x:=0;$ }%
}}}
\put(4873,-324){\makebox(0,0)[lb]{\smash{\SetFigFont{6}{7.2}{\rmdefault}{\mddefault}{\updefault}{\color[rgb]{0,0,0}$q_3$}%
}}}
\put(3022,-890){\makebox(0,0)[lb]{\smash{\SetFigFont{6}{7.2}{\rmdefault}{\mddefault}{\updefault}{\color[rgb]{0,0,0}$L:=0;$ }%
}}}
\put(227,-475){\makebox(0,0)[lb]{\smash{\SetFigFont{6}{7.2}{\rmdefault}{\mddefault}{\updefault}{\color[rgb]{0,0,0}$\dot{x}\in[4/5,1]$}%
}}}
\put(1435,-475){\makebox(0,0)[lb]{\smash{\SetFigFont{6}{7.2}{\rmdefault}{\mddefault}{\updefault}{\color[rgb]{0,0,0}$\dot{x}\in[4/5,1]$}%
}}}
\put(2946,-475){\makebox(0,0)[lb]{\smash{\SetFigFont{6}{7.2}{\rmdefault}{\mddefault}{\updefault}{\color[rgb]{0,0,0}$\dot{x}\in[4/5,1]$}%
}}}
\put(4608,-475){\makebox(0,0)[lb]{\smash{\SetFigFont{6}{7.2}{\rmdefault}{\mddefault}{\updefault}{\color[rgb]{0,0,0}$\dot{x}\in[4/5,1]$}%
}}}
\end{picture}
\end{center}
\caption{Examples of HRD}
\label{fig.onehrd}
\end{figure}

We decide to follow an approach similar to the one adopted in \cite{Wang03}.
That is, our decision atoms will be LH-expressions while our BDD arcs will be labeled
with LH-upperbounds.
A node labeled with decision atom $\sum_ia_ix_i$ together with a corresponding
outgoing arc label $(\sim, c)$ constitute the LH-constraint of $\sum_ia_ix_i\sim c$.
A root-to-terminal path in an HRD thus represents the conjunction 
of constituent LH-constraints along the path.
Figure~\ref{fig.onehrd}(a) is an example of our proposed BDD-like data-strucutre for
the concave space of
\begin{center}
$(x_2-x_3\leq -5/7 \vee -5A-2x_2+10x_3\leq 48/7)\wedge A-x_2+10x_3 < 9$
\end{center}
assuming that $-5A-2x_2+10x_3$ precedes $x_2-x_3$ (in symbols $-5A-2x_2+10x_3\prec x_2-x_3$)
and $x_2-x_3$ precedes $A-x_2+10x_3$ in the given evaluation ordering.
In this example, the system variables are $A,x_2,x_3$ while the decision atoms
are $x_2-x_3$, $-5A-2x_2+10x_3$, and $A-x_2+10x_3$.

{\definition HRD (Hybrid-Restriction Diagram)}
Given dense variable set $X=\{x_1,\ldots,x_n\}$ and
an evaluation ordering $\prec$ among normalized LH-expressions of $X$,
an {\em HRD} is either $\true$ or
a tuple $(v,(\beta_1,D_1),\ldots,(\beta_m,D_m))$
such that
\begin{list1}
\item $v$ is a normalized LH-expression;
\item for each $1\leq i\leq m$, $\beta_i$ is an LH-upperbound s.t.
	$(<,\infty)\neq \beta_1\sqsubset \beta_2\sqsubset \ldots\sqsubset \beta_m$; and
\item for each $1\leq i\leq m$, $D_i$ is an HRD such that if $D_i=(v_i,\ldots)$,
	then $v\prec v_i$.
\end{list1}
For completeness, we use "$()$" to represent the HRD for $\false$.  
\qed

In our algorithms, $\false$ does not participate in comparison 
of evaluation orderings among decision atoms.  
Also, note that in figure~\ref{fig.onehrd}, for each arc label $(\sim,c)$, we simply put down
$\sim c$ for convenience.
Note that an HRD records a set of convex polyhedra and each root-leaf path represents
such a convex polyhedron.

\section{Three dense orderings among decision atoms\label{sec.vorderings}}

In the definition of a dense-ordering among decision atoms (i.e., LH-expressions), 
special care must be taken to facilitate efficient manipulation of HRDs.
Here we use the experience reported in \cite{Wang03} and
present three criteria in designing the orderings among LH-expressions.
The three criteria are presented in sequence proportional to their respective importances.

First, it is desirable to place a pair of converse LH-expressions next to one another
so that simple inconsistencies can be easily detected.
That is, LH-expressions $\sum_ia_ix_i$ and $\sum_i-a_ix_i$ are better placed next to one another
in the ordering.
For example, with this arrangement, the inconsistency of $-x_1+3x_2\leq -5\wedge x_1-3x_2<0$
can be checked by comparing adjacent nodes in HRD paths.
To fulfill this requirement,
when comparing the precedence between LH-expressions in a given ordering,
we shall first toggle the signs of coefficients of an LH-expression
if its first nonzero coefficient is positive.
If two LH-expressions are identical after the necessary toggling,
then we compare the signs of their
first nonzero coefficients to decide the precedence between the two.

With the requirement mentioned in the last paragraph, from now on, we shall only
focus on the orderings among LH-expressions whose first nonzero coefficients are negative.

Secondly, according to past experience reported in the literature,
it is important to place strongly correlated LH-expressions close together in
the evaluation orderings.
Usually, instead of a single global LHA,
we are given a set of communicating LHAs, each representing a process.
Thus it is desirable to
place LH-expressions for the same process close to each other in the orderings.
Our second important criterion respects this experience.
Given a system with $m$ processes with respective local dense variables,
we shall partition the LH-expressions into $m+1$ groups: $G_0,G_1,\ldots,G_m$.
$G_0$ contains all LH-expressions without local variables
(i.e., coefficients for local variables are all zero).
For each $p>0$, $G_p$ contains all LH-expressions with a nonzero coefficient
for a local variable of process $p$ and only zero coefficients for
local variables of processes $p+1,\ldots, m$.
Then our second criterion requires that
for all $0\leq p<m$, LH-expressions in $G_p$ precede those in $G_{p+1},\ldots,G_m$.

If the precedence between two LH-expressions cannot be determined
with the two above-mentioned criteria, 
then the following third criterion comes to play.
This one is a challenge since for each of $G_0,\ldots,G_m$ can be of infinite size.
Traditionally, BDD-like data-structures have been used with finite decision atom domains.
But now we need to find a way to determine the precedence among infinite number 
of LH-expressions (our decision atoms in HRD).
For this purpose, we invent to use dense-orderings among LH-expressions.
We shall present three such orderings in the following.
Sometimes it is difficult to predict which orderings are better suitable for
what kind of verification tasks.
In section~\ref{sec.experiments}, we shall report experiments
with these orderings.\\[2mm]
\noindent {\bf Dictionary ordering:}
We can represent each LH-expression as a string, assuming that
the ordering among $x_1,\ldots,x_n$ is fixed and no blanks are used in the string.
Then we can use dictionary ordering and ASCII ordering to decide the precedence
among LH-expressions.
For the LH-expressions in figure~\ref{fig.onehrd},
we then have $-5A-2x_2+10x_3\prec A-x_2+10x_3 \prec x_2-x_3$ since
'$-$' precedes '$A$' and '$A$' precedes '$x$' in ASCII.
The corresponding HRD in dictionary ordering is in figure~\ref{fig.onehrd}(c).
One interesting feature of this ordering is that it has the potential to
be extended to nonlinear hybrid constraints.
For example, we may say $\cos(x_1)+x_2^3\prec x_2^2-x_2x_3$ in dictionary ordering
since '$c$' precedes '$x$' in ASCII.\\[2mm]
\noindent {\bf Coefficient ordering:}
Assume that the ordering of the dense variables is fixed as $x_1,\ldots,x_m$.
In this ordering, the precedence between two LH-expressions is
determined by iteratively comparing the coefficients of dense variables
$x_1,\ldots,x_n$ in sequence.
For the LH-expressions in figure~\ref{fig.onehrd}, we then have
$-5A-2x_2+10x_3\prec x_2-x_3\prec A-x_2+10x_3$
The HRD in this ordering is in figure~\ref{fig.onehrd}(a).\\[2mm]
\noindent {\bf Magnitude ordering:}
This ordering is similar to the last one.
Instead of comparing coefficients, we compare the absolute values of coefficients.
We iteratively
\begin{list1}
\item first compare the absolute values of coefficients of $x_i$, and
\item if they are equal, then compare the signs of coefficients of $x_i$.
\end{list1}
For the LH-expressions in figure~\ref{fig.onehrd}, we then have
$x_2-x_3\prec A-x_2+10x_3 \prec -5A-2x_2+10x_3$ in this magnitude ordering.
The HRD in this ordering is in figure~\ref{fig.onehrd}(b).

\section{Set-oriented operations \label{sec.set}}

Please be reminded that an HRD records a set of convex polyhedra.
For convenience of discussion, given an HRD, we may just represent
it as the set of covnex polyhedra recorded in it.
Definitions of set-union ($\cup$), set-intersection ($\cap$), and set-exclusion ($-$)
of two convex polyhedra sets respectively represented by two HRDs
are straightforward.
For example, given HRDs $D_1:\{\zeta_1, \zeta_2\}$
and $D_2:\{\zeta_2,\zeta_3\}$,
$D_1\cap D_2$ is the HRD for $\{\zeta_2\}$;
$D_1\cup D_2$ is for $\{\zeta_1,\zeta_2,\zeta_3\}$; and
$D_1- D_2$ is for $\{\zeta_1\}$.
The complexities of the three manipulations are all $O(|D_1|\cdot|D_2|)$.


Given two convex polyhedra $\zeta_1$ and $\zeta_2$,
$\zeta_1\sqcap \zeta_2$ is a new convex polyhdron representing the
space-intersection of $\zeta_1$ and $\zeta_2$.
Formally speaking, for decision atom $\sum_ia_ix_i$, 
$\zeta_1\sqcap\zeta_2(\sum_ia_ix_i)=\zeta_1(\sum_ia_ix_i)$
if $\zeta_1(\sum_ia_ix_i)\sqsubset \zeta_2(\sum_ia_ix_i)$;
or $\zeta_2(\sum_ia_ix_i)$ otherwise.
Space-intersection ($\sqcap$) of two HRDs $D_1$ and $D_2$,
in symbols $D_1\sqcap D_2$, is a new HRD for
$\{\zeta_1\sqcap\zeta_2\mid \zeta_1\in D_1; \zeta_2\in D_2\}$.

Given an evaluation ordering, we can write HRD-manipulation algorithms
pretty much as usual\cite{BCMDH90,Bry86,Wang00a,Wang03}.
For convenience of presentation, we may repersent an HRD 
$(u,(\beta_1,B_1),\ldots,(\beta_n,B_n))$
symbolically as 
$(u,(\beta_i,B_i)_{1\leq i\leq n})$.
A union operation $\cup(B,D)$ can then be implemented as follows.
\procbegin
\noindent
set $\Psi$; /* database for the recording of already-processed cases */ \\
\noindent $\cup(B,D)$ \{
\hst if $B=\false$, return $D$; else if $D=\false$, return $B$; 
\hst $\Psi:=\emptyset$;
    return $\mbox{\tt rec$\cup$}(B,D)$;  
\\\}
\vspace*{2mm}
\\
$\mbox{\tt rec$\cup$}(B, D)$ where
$B=(u,(\beta_i,B_i)_{1\leq i\leq n})$,
$D=(v,(\alpha_j,D_j)_{1\leq j\leq m})$
\{
\hst  if $B$ is $\true$ or $D$ is $\true$, return $\true$;
        else if $\exists F,(B,D,F)\in \Psi$, return $F$; \dotfill (1)
\hst  else if $u\prec v$,
	construct $F:=(u, (\beta_i, \mbox{\tt rec$\cup$}(B_i,D))_{1\leq i\leq n})$;
\hst  else if $v\prec u$,
	construct $F:=(v, (\alpha_j, \mbox{\tt rec$\cup$}(B,D_j))_{1\leq j\leq m})$;
\hst  else \{
\hstt	$i:=n; j:= m; F:=\false;$
\hstt	while $i\geq 1$ and $j\geq 1$, do \{
\hsttt		if $\beta_i=\alpha_j$, \{
\hstttt			$\dot{B}:= \dot{B}\cup B_i; \dot{D}:= \dot{D}\cup D_j;$
			$F:= F\cup (u, (\beta_i, \mbox{\tt rec$\cup$}(B_i,D_j)))$;
			$i--; j--;$
\hsttt		\}
\hsttt		else if $\beta_i\sqsubset \alpha_j$, \{
			$F:= F\cup (u, (\alpha_j,D_j))$;
			$j:= j-1;$
		\}
\hsttt		else if $\alpha_j\sqsubset\beta_i$, \{
			$F:= F\cup (u, (\beta_i, B_i))$;
			$i:= i-1;$
		\}
\hstt	\}
\hstt	if $i\geq 1$, $F:= F\cup (u, (\beta_1, B_1)_{1\leq h\leq i})$;
\hstt	if $j\geq 1$, $F:= F\cup (u, (\alpha_1, D_1)_{1\leq k\leq j})$;
\hst  \}
\hst  $\Psi:=\Psi\cup\{(B,D,F)\}$;
	return $F$;\dotfill (2)
\\\}
\procend

Note that in statement (1), we take advantage of the data-sharing capability
of HRDs so that we do not process the same substructure twice.
The set of $\Psi$ is maintained in statement (2).
The algorithms for $\cap$ and $-$ are pretty much the same.
The one for space intersection is much more involved and is not discussed here
due to page-limit.

\section{HRD+BDD \label{sec.hrd.bdd}}

As reported in the experiment with CRD (Clock-Restriction Diagram)\cite{Wang03},
significant performance improvement can be obtained if
an integrated BDD-like data-structure for both dense constraints and discrete
constraints is used instead of separate data-structure 
for them.
It is also possible to combine HRD and BDD into one data-structure
for fully symbolic manipulation. 
Since HRD only has one sink node: $\true$,
it is more compatible with BDD without FALSE terminal node which
is more space-efficient than ordinary BDD.
There are two things we need to take care of in this combination.
The first is about the interpretation of default values of decision atoms.
In BDD, when we find a decision atom is missing during valuating variables along a path,
the atom's value can be interpreted as either TRUE or FALSE.
But in HRD, when we find a decision atom $\sum_i a_ix_i$ is missing along a path,
then the constraint is interpreted as $\sum_i a_ix_i <\infty$.

The second is about the interpretation of HRD manipulations to BDD decision atoms.
Straightforwardly,
``$\cup$'' and ``$\cap$'' on BDD decision atoms are respectively interpreted
as ``$\vee$'' and ``$\wedge$'' on BDD decision atoms.
$D_1-D_2$ on BDD decision atoms is interpreted as $D_1\wedge \neg D_2$
when the root variable of either $D_1$ or $D_2$ is Boolean.
For $D_1\sqcap D_2$, the manipulation acts as ``$\wedge$'' when either
of the root are labeled with BDD decision atoms.
Due to page-limit, we shall omit the proof for the soundness of such an interpretation.
From now on, we shall call it HRD+BDD a combination structure of HRD and BDD.

Finally, it is also important to define the evaluation orderings 
between BDD decision atoms and HRD decision atoms.
Due to page-limit, we shall adopt the wisdom reported in \cite{Wang03} and place
BDD decision atoms and HRD decision atoms that are strongly related 
to the same process close to each other.

\section{Weakest preconditon calculation and symbolic parametric analysis\label{sec.wpc}}

Our tool {\tt red} runs in backward reachability analysis by default.
Due to page-limit, we shall only present the algorithm in symbolic fashion without
details.
Suppose we are given an LHA $A=\langle X, Q, I, \mu, \gamma, E, \tau, \pi\rangle$.
There are two basic procedures in this analysis procedure.
The first, $\mbox{\tt xtion}(D,e)$, computes the weakest precondition from
state-space represented by HRD $D$ through discrete transition $e=(q,q')$.
Assume that the dense variables that get assigned in $e$ are $y_1,\ldots,y_k$
and there is no variable that gets assigned twice in $e$.
The characterization of $\mbox{\tt xtion}(D,e)$ is
\begin{center}
$\mu(q)\sqcap\tau(e)\sqcap\exists y_1\ldots \exists y_k
(D\sqcap \bigsqcap_{1\leq i\leq k}y_i\in \pi(e,y_i))$\footnote{
$y\in [d,d']\equiv d\leq y\leq d'$.
$y\in (d,d']\equiv d< y\leq d'$.
$y\in [d,d')\equiv d\leq y< d'$.
$y\in (d,d')\equiv d< y< d'$.
}
\end{center}

Assume that $\mbox{\tt delta\_exp}(D)$ is the same as $D$ except that all dense variables
$x$ are replaced by $x+\delta_x$ respectively.
Here $\delta_x$ represents the value-change of variable $x$ in time-passage.
For example,
$\mbox{\tt delta\_exp}(2x_1-3x_2\leq 3/5)=2x_1+2\delta_{x_1}-3x_2-3\delta_{x_2}\leq 3/5$.
Intuitively, when $x$ represents the value of variable $x$ in the weakest precondition of
time passage,
then $x+\delta_x$ is the value of $x$ in the postcondition of the time-passage.

The second basic procedure, $\mbox{\tt time}(D, q)$, computes the weakest precondition
from $D$ through time passage in mode $q$.
It is characterized as
\begin{center}
$\mu(q)\sqcap\exists \delta_{x_1}\exists \delta_{x_2} \ldots \exists \delta_{x_n} \exists \delta
\left(
\begin{array}{ll}
	& \delta \geq 0 \sqcap \mbox{\tt delta\_exp}(D) \\
\sqcap 	& \bigsqcap_{1\leq i\leq n; \gamma(q,x_i)=\langle d_i,d'_i\rangle}
		\delta_{x_i}\in \langle d_i\delta, d'_i\delta\rangle
\end{array}\right)$
\end{center}
One basic building block of both $\mbox{\tt xtion}()$ and $\mbox{\tt time}()$
is for the evaluation of $\exists x (D(x))$.
We implement this basic operation with the following symbolic procedure.
\begin{center}
$\exists x(D(x))\equiv \mbox{\tt var\_del}(\mbox{\tt xtivity}(D, x),\{x\})$.
\end{center}
Procedure $\mbox{\tt var\_del}(D, X')$ eliminates all constraints in
$D$ involving variables in set $X'$.
Procedure $\mbox{\tt xtivity}(D,x)$ adds to a path every constraint that can be
transitively deduced from two peer constraints involving $x$ in the same path in $D$.
The algorithm of $\mbox{\tt xtivity}()$ is in table~\ref{tab.xtivity}.
\begin{table}[t]
\hrule
\noindent
\mbox{\tt set} $R$, $S$; \\
$\mbox{\tt xtivity}(D, x)$ \{ $R:=\emptyset$; return $\mbox{\tt rec\_xtivity}(D)$; \}\\
$\mbox{\tt rec\_xtivity}(D)$ \{
\hst	if $D$ is $\true$ or $\false$, return $D$;
	else if $\exists (D,D')\in R$, return $D'$;
\hst	else /* assume $D=(a x+\epsilon, (\beta_1,D_1),\ldots, (\beta_m,D_m))$ */ \{
\hstt		$S:=\emptyset;
		D':=\bigcup_{1\leq i\leq m}
			a x+\epsilon \beta_i\sqcap \mbox{\tt rec\_xtivity\_given}
				(D_i,ax+\epsilon, \beta_i)$;
\hstt		$R:=R\cup\{(D,D')\}$; return $D'$;
\hst	\}
\\\}\\
$\mbox{\tt rec\_xtivity\_given}(D, ax+\epsilon,\beta)$ \{
\hst	if $D$ is $\true$ or $\false$, return $D$;
	else if $\exists (D,D')\in S$, return $D'$; \dotfill (3)
\hst	else /* assume $D=(b x+\epsilon', (\beta_1,D_1),\ldots, (\beta_m,D_m))$ */ \{
\hstt		if $ab<0$,
\hsttt			$D':=\bigcup_{1\leq i\leq m}
			\left(\begin{array}{ll}
				& b x+\epsilon' \beta_i\sqcap\mbox{\tt rec\_xtivity\_given}
				(D_i,ax+\epsilon, \beta)\\
			\sqcap	&|b|\epsilon/\gcd(a,b) + |a|\epsilon'/\gcd(a,b)((|b|\beta+|a|\beta_i)/\gcd(a,b))
			\end{array}\right)$;
\hstt		else $D':=\bigcup_{1\leq i\leq m}
			b x+\epsilon' \beta_i\sqcap
			\mbox{\tt rec\_xtivity\_given}
				(D_i,ax+\epsilon, \beta)$;
\hstt		$S:=S\cup\{(D,D')\}$; return $D'$; \dotfill (4)
\hst	\}
\\\}
\hrule
$(|b|\beta+|a|\beta_i)/\gcd(a,b)$ is a shorthand for the new upperbound obtained from the
xtivity of $ax+\epsilon \beta$ and $bx+\epsilon' \beta_i$.
\caption{Algorithm for xtivity()}
\label{tab.xtivity}
\end{table}
Thus we preserve all constraints transitively deducible from
a dense variable before it is eliminated from a predicate.
This guarantees that no information will be unintentionally lost after the variable elimination.

Note that in our algorithm, we do not enumerate all paths in HRD to carry
out this least fixpoint evaluation.
Instead, in statement (3), our algorithm follows the traditional BDD programming style which takes
advantage of the data-sharing capability of BDD-like data-structures.
Thus our algorithm does not explode due to the combinatorial complexity of path counts
in HRD.
This can be justified by the performance of our implementation reported in section~\ref{sec.experiments}.

Assume that the unsafe state is in mode $q_f$.
With the two basic procedures, then
the backward reachable state-space from the risk state $\neg\eta$ (represented as an HRD)
can be characterized by
\begin{center}
$\mbox{\tt lfp} Z.(\mbox{\tt time}(\neg\eta,q_f)\cup
	\bigcup_{e=(q,q')\in E}\mbox{\tt time}(\mbox{\tt xtion}(Z,e), q))$
\end{center}
Here $\mbox{\tt lfp} Z.F(Z)$ is the least fixpoint of function $F()$ and is
very commonly used in the reachable state-space representation of discrete and
dense-time systems.
After the fixpoint is successfully constructed, we conjunct it with the
initial condition and then eliminate all variables except those static parameters
(formally speaking, projecting the reachable state-space representations to
the dimensions of the static parameters).
Suppose the set of static dense parameters is $H$.
The characterization of unsafe parameter valuatons is thus
\begin{center}
$\mbox{\tt var\_del}(I\sqcap\mbox{\tt lfp} Z.(\mbox{\tt time}(\neg\eta,q_f)\cup
	\bigcup_{e=(q,q')\in E}\mbox{\tt time}(\mbox{\tt xtion}(Z,e), q)),
	X-H)$
\end{center}
The set of parametric solutions is characterized by the complement of this final result.

\section{Normalization \label{sec.norm}}

There can be infinitely many LH-constraint sets that
represent a given convex polyhedron.
An LH-constraint in such a representation can also be {\em redundant} in that
a no less restrictive upperbound can be derived for its LH-expression
from peer LH-constraints in the same representation.
To control the redundancy caused by recording many LH-constraint sets for the
same convex polyhedron, representations of convex polyhedra have to be normalized.
Due to page-limit, we shall skip much details in this regard.
We emphasize that much of our implementation effort has been spent in this regard.
We use a two-phase normalization procedure in each iteration of the least fixpoint
evaluation.
\begin{list1}
\item[\bf Step I, for subsumed polyhedra elimination]:
	This step eliminates those convex polyhedra contained by a peer convex polyhedron
	in the HRD for the reachable state-space.
	First, we collect the LH-expressions that occur in the current reachable state-space
	HRD and call them {\em proof-obligations}.
	Then we try to derive the tightest constraints for these proof-obligations
	along each HRD paths of the reachable state-space representation.
	Then we eliminate those paths which is subsumed by other paths.
	The subsumption can be determined by pairwise comparison of all LH-constraints
	along two paths.
\item[\bf Step II, for redundant constraint elimination]:
	Along each path, we combinatorially use up to four constraints to check for
	the redundancy of peer constraints in the same path and eliminate them if they are
	found redundant.
\end{list1}
Again, our algorithm does not enumerate paths in HRD.
Instead, it takes advantage of data-sharing capability of HRD for
efficient processing.

\section{Pruning strategy based on parameter space construction (PSPSC)
	\label{sec.pspsc}}

We have also experimented with techniques to improve the efficiency of
parametric analysis.
One such technique, called {\em PSPSC}, is avoiding new state-space exploration 
if the exploration does not contribute to new parametric solutions.
A constraint is {\em static} iff all its dense variables are static parameters.
Static constraints do not change their truth values.
Once a static constraint is derived in a convex polyhedron, its truth value
will be honored in all weakest preconditions derived from this convex polyhedron.
All states backwardly reachable from a convex polyhedron must
also satisfy the static constraints required in the polyhedron.
Thus if we know that static parameter valuation $\cal H$ is already in
the parametric solution space, then we really do not need to explore those states
whose parameter valuations fall in $H$.

With PSPSC, our new parametric analysis procedure is shown in table~\ref{tab.PSPSC}.
\begin{table}
\hrule
\noindent
$\mbox{\tt PSA\_with\_PSPSC}(A,\eta)$ \{
\hst	$\bar{D}:=\mbox{\tt time}(\neg\eta,q_f); D:=\false;
	P:=\mbox{\tt var\_del}(D, X-H);$
\hst	while $\bar{D}\neq \false$, do \{
\hstt		$D:=D\cup \bar{D}$;
\hstt		$\bar{D}:=\bigcup_{e=(q,q')\in E}\mbox{\tt time}(\mbox{\tt xtion}(\bar{D},e), q)$;
\hstt		$\bar{D}:=\bar{D}\sqcap (\neg P)\sqcap (\neg D)$;\dotfill (5)
\hstt		$P:=P\cup \mbox{\tt var\_del}(I\sqcap \bar{D}, X-H);$
\hst	\}
\hst	return $\neg P$;
\\\}
\hrule
\caption{Procedure for parametric safety analysis with PSPSC}
\label{tab.PSPSC}
\end{table}
In the procedure, we use varaible $P$ to symbolically accumulate 
the parametric evaluations leading to the risk states 
in the least fixpoint iterations.
In statement (5), we check and eliminate in $\bar{D}$ those state descriptions which cannot
possibly contribute to new parametric evaluations by conjuncting $\bar{D}$ with $\neg P$.

One nice feature of PSPSC is that it does not sacrifice the precision of our parametric 
analysis. 

{\lemma \label{lemma.pspsc} 
$\cal H$ is a {\em parametric solution} to $A$ and $\eta$ iff
$\cal H$ satisfies the return result of 
$\mbox{\tt PSA\_with\_PSPSC}(A,\eta)$.
}
\\\pf 
Details omitted due to page-limit.  
The basic idea is that the intersection at line (5) in table~\ref{tab.PSPSC} 
only stops the further exploration of those states that do not contribute
to new parameter-spaces.  
Those parameter-spaces pruned in line (5) do not contribute 
because they are already contained in the known parameter constraints $P$ 
and along each exploration path, the parameter constraints only get restricter.  
\qed 

As mentioned in the proof sketch, PSPSC can help in pruning the space of exploration 
in big chunks.  
But in the worst case, PSPSC does not guarantee the exploration will terminate.  
In section~\ref{sec.experiments}, we shall report the performance of this technique.
Especially, for one benchmark, the state-space exploration cannot converge 
without PSPSC.

\section{Implementation and experiments\label{sec.experiments}}

We have implemented our ideas in our tool {\tt red}
which has been previously reported in \cite{Wang00a,Wang00b,Wang01a,Wang01b,Wang03} for
the verification of timed automata based on BDD-like data-structures.
{\tt red} version 5.0 supports full TCTL model-checking/simulation with graphical user-interface.
Coverage estimation techniques for dense-time state-spaces has also been reported\cite{WHY03b}.

\subsection{Comparison with HyTech 2.4.5}

We have also carried out experiments to compare various ideas mentioned in this work.
In addition, we have also compared with HyTech 2.4.5\cite{HHWt95},
which is the best known and most
popular tool for the verification of LHA due to its pioneering importance.
The following three benchmark series are all adapted from HyTech benchmark repository.
\begin{list1}
\item {\em Fischer's mutual exclusion algorithm.}
	This is one of the classic benchmarks.
	There are two static parameters $A$ and $B$, $m$ processes, and
	one local clock for each process.
	The first process has a local clock with rate in $[4/5,1]$ while
	all other processes have local clocks with rates in $[1,11/10]$.
	The algorithm may violate the mutual exclusion property when $-A\leq 0\wedge -11A+8B\leq 0$.
\item {\em General railroad crossing benchmarks.}
	There is a static parameter {\tt CUTOFF}, a gate-process, a controller-process,
	and $m$ train-processes.
	The local dense variable of the gate-process models the angle of the gate and
	has rates in $[0,0]$, $[-10,-9]$, and $[9,10]$ depending on which modes the gate-process
	is in.
	The controller process does not use clocks.
	Each train-process uses a local clock with rate in $[1,1]$.
	The system may not lower the gate in time for a crossing train when
	$20\leq \mbox{\tt CUTOFF}\leq 40$.
\item {\em Nuclear reactor controller.}
	There are $m$ rod-processes and one controller process.
	Each process has a clock with rate in $[1,1]$.
	A rod just-moved out of the heavy water must stay out of water for at least $T$ (a static
	parameter) time units.
	The timing constants used in the benchmarks are $58/10,59/10,16$, and $161/10$.
	The controller may miss the timing-constraints for the rods if
	$-T\leq -(109m-29)/5$.
\item{\em CSMA/CD.}
	This is modified from \cite{Yovine97}.
	The two timing constants $A$ and $B$, set to 26 and 52 respectively, are now
	treated as static parameters to be analyzed.
	We do require that $B\geq 52$.
	Basically, this is the ethernet bus arbitration protocol with
	the idea of collision-and-retry.
	The biggest timing constant used is 808.
	We want to verify that mutual exclusion after bus-contending period can
	be violated if
	$A>0\wedge B\geq 52\wedge B\leq 808\wedge B<2A$.
\end{list1}

In our experiment, we compare performance in both forward and backward reachability analyses.
The performance data of HyTech 2.4.5 and {\tt red} 5.0 with dictionary ordering (no PSPSC),
coefficient ordering (no PSPSC), magnitude ordering (PSPSC), and
coefficient ordering with PSPSC is reported in table~\ref{tab.perf.bck} (for backward
analysis) and table~\ref{tab.perf.fwd} (for forward analysis).
\begin{table}[t]
\begin{center}
{\scriptsize
\renewcommand{\tabcolsep}{0pt}
\begin{tabular}{|l|c||r|r|r|r|r|}
\hline
benchmarks	& concurrency	& \multicolumn{1}{c|}{HyTech}	& \multicolumn{4}{c|}{{\tt red} 5.0 (backward)} \\ \cline{4-7}
		&		& \multicolumn{1}{c|}{2.4.5}	& \multicolumn{1}{c|}{dictionary}	& \multicolumn{1}{c|}{coefficient} & \multicolumn{1}{c|}{magnitude}	& \multicolumn{1}{c|}{coefficient}	\\ \cline{4-7}
		&		& (backward)	& \multicolumn{3}{c|}{no PSPSC}						& \multicolumn{1}{c|}{PSPSC} \\\hline \hline
Fischer's 	& 2 procs	& 0.23s		& 0.10s/17k	& 0.11s/17k	& 0.11s/17k	& 0.07s/16k \\ \cline{2-7}
mutual 		& 3 procs	& 2.40s		& 1.83s/81k	& 1.75s/74k	& 1.23s/59k	& 0.70s/44k \\ \cline{2-7}
exclusion	& 4 procs	& 28.04s	& 20.29s/320k	& 23.85s/269k	& 12.38s/215k	& 5.14s/163k \\ \cline{2-7}
($m$ 		& 5 procs	& O/M		& 278.8s/1420k	& 354.1s/1149k	& 162.0s/1034k	& 31.36s/474k \\ \cline{2-7}
		& 6 procs	& O/M		& 2846s/5848k	& 9923s/8796k	& 1485s/4000k	& 168.6s/1170k \\\hline\hline
general		& 2 trains	& O/M		& 0.79s/103k	& 0.68s/101k	& 0.68s/101k	& 0.76s/94k \\ \cline{2-7}
railroad 	& 3 trains	& O/M		& 11.48s/806k	& 8.85s/616k	& 8.84s/616k	& 11.48s/530k \\ \cline{2-7}
crossing	& 4 trains	& O/M 		& 248.5s/6046k	& 184.9s/4249k	& 186.1s/4249k	& 252.5s/2820k \\ \cline{2-7}
		& 5 trains	& O/M 		& 6095s/37093k	& 4883s/25841k	& 4900s/25841k	& 6527s/19234k \\ \hline\hline
reactor		& 2 rods	& 0.056s	& 0.08s/19k	& 0.07s/19k	& 0.06s/19k	& 0.05s/15k \\ \cline{2-7}
($m$ rods)	& 3 rods	& 0.33s		& 0.41s/51k	& 0.38s/52k	& 0.37s/52k	& 0.22s/41k \\ \cline{2-7}
		& 4 rods	& 2.61s		& 3.10s/187k	& 2.69s/186k	& 2.71s/186k	& 1.42s/155k \\ \cline{2-7}
		& 5 rods	& 31.29s	& 41.47s/1042k	& 37.03s/1039k	& 36.89s/1039k	& 18.67s/884k \\ \cline{2-7}
		& 6 rods	& 647.8s	& 951.5s/8228k	& 866.9s/8191k	& 839.3s/8191k	& 461.8s/6941k \\ \hline\hline
CSMA/CD		& 2 senders	& O/M		& 0.98s/42k$^*$	& 1.47s/125k	& 0.57s/34k	& 0.56s/33k \\ \cline{2-7}
($m$ senders)	& 3 senders	& O/M		& O/M		&5076s/2407k$^*$& 121.5s/807k	& 0.66s/105k \\ \cline{2-7}
		& 4 senders	& O/M		& O/M		& O/M		& O/M	 	& 2.47s/378k \\ \cline{2-7}
		& 5 senders	& O/M		& O/M		& O/M		& O/M		& 9.77s/1192k \\ \cline{2-7}
		& 6 senders	& O/M		& O/M		& O/M		& O/M		& 40.58s/3513k \\ \hline
\end{tabular}
}
\\
data collected on a Pentium 4M 1.6GHz with 256MB memory running LINUX; \\
s: seconds; k: kilobytes of memory in data-structure; O/M: Out of memory; 
\end{center}
\caption{Comparison in backward analysis with HyTech w.r.t. number of processes}
\label{tab.perf.bck}
\end{table}
\begin{table}[t]
\begin{center}
{\scriptsize
\renewcommand{\tabcolsep}{0pt}
\begin{tabular}{|l|c||r|r|r|r|r|}
\hline
benchmarks	& concurrency	& \multicolumn{1}{c|}{HyTech}	& \multicolumn{4}{c|}{{\tt red} 5.0 (forward)} \\ \cline{4-7}
		&		& \multicolumn{1}{c|}{2.4.5}	& \multicolumn{1}{c|}{dictionary}	& \multicolumn{1}{c|}{coefficient} & \multicolumn{1}{c|}{magnitude}	& \multicolumn{1}{c|}{coefficient}	\\ \cline{4-7}
		&		& (forward)	& \multicolumn{3}{c|}{no PSPSC}						& \multicolumn{1}{c|}{PSPSC} \\\hline \hline
Fischer's 	& 2 procs	& 0.34s		& 0.10s/20k	& 0.10s/20k	& 0.10s/19k	& 0.08s/18k \\ \cline{2-7}
mutual 		& 3 procs	& 37.89s	& O/M		& 22.10s/561k	& 19.18s/654k	& 5.59s/538k \\\hline\hline
general		& 2 trains	& 3.29s		& 2.29s/192k	& 1.41s/95k	& 1.43s/95k	& 0.44s/84k \\ \cline{2-7}
railroad 	& 3 trains	& O/M		& O/M	& O/M		& O/M		& 6.35s/418k \\ \hline\hline
reactor		& 2 rods	& O/M		& O/M	& O/M	 	& O/M	 	& O/M \\ \hline\hline
CSMA/CD		& 2 senders	& 0.19s		& 0.19s/29k	& 0.17s/29k	& 0.17s/29k	& 0.25s/33k \\ \cline{2-7}
($m$ senders)	& 3 senders	& 2.63s		& 1.81s/102k	& 1.64s/101k	& 1.62s/101k	& 2.61s/106k \\ \cline{2-7}
		& 4 senders	& 68.75s	& 20.07s/370k	& 17.49s/378k	& 17.52s/378k	& 27.03s/378k \\ \cline{2-7}
		& 5 senders	& O/M		& 268.0s/1905k	& 240.3s/1906k	& 242.2s/1906k	& 331.9s/1910k \\ \cline{2-7}
		& 6 senders	& O/M		& 3889s/11725k	& 3123s/11525k 	& 3155s/11525k	& 4163s/11552k \\ \hline
\end{tabular}
}
\\
data collected on a Pentium 4M 1.6GHz with 256MB memory running LINUX; \\
s: seconds; k: kilobytes of memory in data-structure; O/M: Out of memory; 
\end{center}
\caption{Comparison in forward analysis with HyTech w.r.t. number of processes}
\label{tab.perf.fwd}
\end{table}
We stop experimenting with higher concurrency when we feel that too much time
(like more than 1 hour) or too much memory (20MB) has been consumed in early fixpoint iterations.
The experiment, although not extensive,
does show signs that HRD-technology (with or without PSPSC) can compete with the technology used in HyTech 2.4.5.
For all the benchmarks, HRD-technology demonstrates better scalability w.r.t. concurrency
complexity.
We believe that the data-sharing capability of HRD, when properly programmed,
is the main reason for the performance advantage in the experiment.

Finally, PSPSC cuts down the time and memory needed for parametric analysis.
Especially, in forward analysis of the general railroad benchmark with three trains, 
without PSPSC, the state-space exploration fails to converge.  
This shows very good promise of this technique.

\subsection{Comparison with TReX 1.3}

Another famous tool for the verification of hybrid systems is TReX\cite{AAB00,ABS01},
which supports the verification of systems with clocks, static parameters,
and lossy channels.
As mentioned in section~\ref{sec.relwork}, the time-progress weakest precondition
(or strongest postcondition in forward analysis) calculation algorithm in red 5.0 is
more complex than the one in TReX.
And TReX now mainly runs in forward analysis.
And TReX also may have tuned its performance for systems with lossy channels.
Thus it can be difficult to compare the performance of TReX with {\tt red} 5.0 directly.
Anyway, we still tried hard and used one week to learn the input language of TReX
and to analyze two benchmarks.
The first is Fischer's protocol with all clocks in the uniform rate of 1.
The second is the Nuclear Reactor Controller.
The performance data is shown in table~\ref{tab.perf.trex} for 
both forward and backward analysis.  
\begin{table}[t]
\begin{center}
{\scriptsize
\renewcommand{\tabcolsep}{0pt}
\begin{tabular}{|l|c||r|r|r||r|r|r|}
\hline
benchmarks	& concurrency	& \multicolumn{3}{c||}{Forward}	& \multicolumn{3}{c|}{Backward} \\ \cline{3-8}
		& 		& TReX 	& \multicolumn{2}{c||}{{\tt red} 5.0 }
				& TReX 	& \multicolumn{2}{c|}{{\tt red} 5.0 }\\ \cline{4-5}\cline{7-8}
		&		& \multicolumn{1}{c|}{1.3} 
				& magnitude
				& coeff.+PSPSC
				& \multicolumn{1}{c|}{1.3} 
				& magnitude
				& coeff.+PSPSC	\\\hline \hline
Fischer's 	& 2 procs	& 1.12s		& 0.07s/17k	& 0.07s/15k	& 8.96s		& 0.08s/13k	& 0.04s/13k \\ \cline{2-8}
mutual 		& 3 procs	& O/M		& 1.86s/137k	& 0.78s/79k	& $>$4197s 	& 0.66s/43k	& 0.49s/43k \\ \cline{2-8}
exclusion	& 4 procs	& O/M		& 197.9s/2714k	& 16.92s/539k	& N/A		& 5.81s/180k 	& 3.58s/158k\\ \cline{2-8}
($m$ 		& 5 procs	& O/M		& $>$1800s	& 752.7s/5254k	& N/A		& 59.27s/945k 	& 24.71s/658k\\ \cline{2-8}
		& 6 procs	& O/M		& N/A		& $>$1800s	& N/A		& 567.1s/4341k 	& 170.3s/2798k \\\hline\hline
reactor		& 2 rods	& O/M		& O/M		& O/M		& N/A		& 0.06s/19k	& 0.05s/15k \\ \cline{2-8}
($m$ rods)	& 3 rods	& O/M		& O/M		& O/M		& N/A		& 0.37s/52k	& 0.22s/41k \\ \cline{2-8}
		& 4 rods	& O/M		& O/M		& O/M		& N/A		& 2.71s/186k	& 1.42s/155k \\ \cline{2-8}
		& 5 rods	& O/M		& O/M		& O/M		& N/A		& 36.89s/1039k	& 18.67s/884k \\ \cline{2-8}
		& 6 rods	& O/M		& O/M		& O/M		& N/A		& 839.3s/8191k	& 461.8s/6941k \\ \hline
\end{tabular}
}
\\
Data for TReX (backward analysis) is collected on a Pentium III 1GHz/900MB running Linux with 
CPU time normalized with factor $1/1.6$. \\
Data for {\tt red} and for TReX (forward analysis) is collected on a Pentium 4M 1.6GHz/256MB running LINUX. \\
s: seconds; k: kilobytes of memory in data-structure; \\
O/M: Out of memory; N/A: not available; \\
\end{center}
\caption{Performance comparison with TReX w.r.t. number of processes}
\label{tab.perf.trex}
\end{table}
Two additional options of {\tt red} 5.0 were chosen: 
coefficient evaluation ordering with PSPSC and 
magnitude evaluation ordering without. 
At this moment, since we do not have the reduce library, which is not free, 
in TReX for backward analysis, TReX team has kindly collected  
TReX's performance in backward analysis for us.  
Although the data set is still small and incomplete, 
but we feel that the HRD-technology shows a lot of promise in the table. 
We believe this can largely be attributed to the data-sharing capability of 
BDD-like data-structures.

\section{Summary \label{sec.conc}}

This work is a first step toward using BDD-technology for the verification of LHAs.
Although the initial experiment data shows good promise, we feel that there are still many
issues worthy of further research to check the pros and cons of HRD-technology.
Especially, we have to admit that we have not developed algorithms to eliminate
general redundant constraints in HRDs.
Our present implementation eliminates redundant LH-constraints that can be deduced by
four peer LH-constraints along the same paths.
We also require that the LH-expression of the redundant LH-constraint
must not precede the LH-expressions of these four peer LH-constraints.
Although our current implementation does perform well against the
benchmarks, we still hope that there is a better way to check redundancy.

Also, subsumption is another challenge.
Straightforward implementation may use the complement of the current reachable state-space
to filter those newly constructed weakest preconditions.
Since the HRD of the current reachable state-space can be huge,
its complement is very expensive to construct and maintain.


\begin{thebibliography}{10}

\bibitem{AAB00} A. Annichini, E. Asarin, A. Bouajjani.
\newblock
	Symbolic Techniques for Parametric Reasoning about Counter and Clock Systems.
\newblock
	CAV'2000, LNCS 1855, Springer-Verlag.

\bibitem{ABKMPR97} Asaraain, Bozga, Kerbrat, Maler, Pnueli, Rasse.
\newblock
	Data-Structures for the Verification of Timed Automata.
\newblock
	Proceedings, HART'97, LNCS 1201.

\bibitem{ABS01} A. Annichini, A. Bouajjani, M. Sighireanu.
\newblock
	TReX: A Tool for Reachability Analysis of Complex Systems.
\newblock
	CAV'2001, LNCS, Springer-Verlag.


\bibitem{ACHH93} R. Alur, C.Courcoubetis, T.A. Henzinger, P.-H. Ho.
\newblock
       Hybrid Automata: an Algorithmic Approach to the Specification
       and Verification of Hybrid Systems.
\newblock
	Proceedings of Workshop
       on Theory of Hybrid Systems, LNCS 736, Springer-Verlag, 1993.

\bibitem{ACHHHNOSS95} R. Alur, C. Courcoubetis, N. Halbwachs, T.A. Henzinger,
	P.-H. Ho, X. Nicollin, A. Olivero, J. Sifakis, S. Yovine.
	The Algorithmic Analysis of Hybrid Systems.
	Theoretical Computer Science 138(1995) 3-34, Elsevier Science B.V.

\bibitem{AD89} R. Alur, D.L. Dill.
\newblock
    Automata for modelling real-time systems.
\newblock
    ICALP' 1990, LNCS 443, Springer-Verlag, pp.322-335.

\bibitem{AHH93} R. Alur, T.A. Henzinger, P.-H. Ho.
	Automatic Symbolic Verification of Embedded Systems.
 	in Proceedings of 1993 IEEE Real-Time System Symposium.

\bibitem{AHV93} R. Alur, T.A. Henzinger, M.Y. Vardi. 
         Parametric Real-Time Reasoning,
         in Proceedings, 25th ACM STOC, pp.~592--601.


\bibitem{Balarin96} F. Balarin.
\newblock
	Approximate Reachability Analysis of Timed Automata.
\newblock
	IEEE RTSS, 1996.

\bibitem{BCMDH90} J.R. Burch, E.M. Clarke, K.L. McMillan, D.L.Dill, L.J. Hwang.
\newblock
	Symbolic Model Checking: $10^{20}$ States and Beyond.
\newblock
	IEEE LICS, 1990.




\bibitem{BLPWW99} G. Behrmann, K.G. Larsen, J. Pearson, C. Weise, Wang Yi.
\newblock
	Efficient Timed Reachability Analysis Using Clock Difference Diagrams.
\newblock
	CAV'99, July, Trento, Italy, LNCS 1633, Springer-Verlag.

\bibitem{Bry86} R.E. Bryant.
\newblock
	Graph-based Algorithms for Boolean Function Manipulation,
\newblock
	IEEE Trans. Comput., C-35(8), 1986.

\bibitem{Dill89} D.L. Dill.
\newblock
	Timing Assumptions and Verification of Finite-state Concurrent
	Systems.
\newblock
	CAV'89, LNCS 407, Springer-Verlag.






\bibitem{HHWt95} T.A. Henzinger, P.-H. Ho, H. Wong-Toi.
\newblock
	HyTech: The Next Generation.
\newblock
 	in Proceedings of 1995 IEEE Real-Time System Symposium.


\bibitem{HNSY92} T.A. Henzinger, X. Nicollin, J. Sifakis, S. Yovine.
\newblock
	Symbolic Model Checking for Real-Time Systems.
\newblock
	IEEE LICS 1992.






\bibitem{MLAH99a} J. Moller, J. Lichtenberg, H.R. Andersen, H. Hulgaard.
\newblock
	Difference Decision Diagrams.
\newblock
	In proceedings of Annual Conference
	of the European Association for Computer Science Logic (CSL), Sept. 1999, Madreid, Spain.

\bibitem{MLAH99b} J. Moller, J. Lichtenberg, H.R. Andersen, H. Hulgaard.
\newblock
	Fully Symbolic Model-Checking of Timed Systems using Difference Decision Diagrams,
\newblock
	In proceedings of Workshop on Symbolic Model-Checking (SMC), July 1999, Trento, Italy.




\bibitem{PL00} P. Pettersson, K.G. Larsen.
\newblock
	UPPAAL2k.
\newblock
	Bulletin of the European Associatoin for Theoretical Computer Science, vol. 70,
	pp.40-44, 2000.


\bibitem{Wang00a} F. Wang.
\newblock
	Efficient Data-Structure for Fully Symbolic Verification
	of Real-Time Software Systems.
\newblock
	TACAS'2000, March, Berlin, Germany; LNCS 1785, Springer-Verlag.

\bibitem{Wang00b} F. Wang.
\newblock
	Region Encoding Diagram for Fully Symbolic Verification of Real-Time Systems.
\newblock
	The 24th COMPSAC, Oct. 2000, Taipei, Taiwan, ROC, IEEE press.

\bibitem{Wang01a} F. Wang.
\newblock
	RED: Model-checker for Timed Automata with Clock-Restriction Diagram.
\newblock
	Workshop on Real-Time Tools, Aug. 2001, Technical Report 2001-014, ISSN 1404-3203, Dept. of Information
	Technology, Uppsala University.

\bibitem{Wang01b} F. Wang.
\newblock
	Symbolic Verification of Complex Real-Time Systems with Clock-Restriction Diagram.
\newblock
	In proceedings of FORTE'2001, Kluwer; August 2001, Cheju Island, Korea.


\bibitem{Wang03} F. Wang.
\newblock
     Efficient Verification of Timed Automata
     with BDD-like Data-Structures, 
\newblock
     to appear in special issue of STTT (Software Tools for Technology Transfer, 
     Springer-Verlag) for VMCAI'2003.  
     The conference version is in proceedings of VMCAI'2003, LNCS 2575, Springer-Verlag.



\bibitem{WHY03b}  F. Wang, G.-D. Hwang, F. Yu.
	Numerical Coverage Estimation for the Symbolic Simulation
	of Real-Time Systems.
	To appear in the proceedings of FORTE'2003, Sept.-Oct. 2003, Berlin, Germany;
	LNCS, Springer-Verlag.

\bibitem{WME93} F. Wang, A. Mok, E.A. Emerson.
 	Symbolic Model-Checking for Distributed Real-Time Systems.
 	In proceedings of 1st FME, April 1993, Denmark; LNCS 670, Springer-Verlag.


\bibitem{WongToi95} H. Wong-Toi.
	Symbolic Approximations for Verifying Real-Time Systems.
	Ph.D. thesis, Stanford University, 1995.


\bibitem{Yovine97} S. Yovine.
\newblock
	Kronos: A Verification Tool for Real-Time Systems.
\newblock
	International Journal of Software Tools for Technology Transfer,
	Vol. 1, Nr. 1/2, October 1997.


\end{thebibliography}

\end{document}